# Cascade Brillouin scattering as a mechanism for photoluminescence from rough surfaces of noble metals

V. Yu. Shishkov,[1,2,3] E. S. Andrianov,[1,2] A. A. Pukhov,[1,2,3] A. P. Vinogradov,[1,2,3] S. N. Orlov,[4] Yu. N. Polivanov,[4] V. I. Fabelinsky,[4] D. N. Kozlov,[4] V. V. Smirnov,[4] and A. A. Lisyansky[5,6]

[1]Dukhov Research Institute of Automatics (VNIIA), 22 Sushchevskaya, Moscow 127055, Russia
[2]Moscow Institute of Physics and Technology, 9 Institutskiy per., Dolgoprudny 141700, Moscow reg., Russia
[3]Institute for Theoretical and Applied Electromagnetics, 13 Izhorskaya, Moscow 125412, Russia
[4]Prokhorov General Physics Institute of the Russian Academy of Sciences, Vavilov str. 38, 119991 Moscow, Russia
[5]Department of Physics, Queens College of the City University of New York, Flushing, New York 11367, USA
[6]The Graduate Center of the City University of New York, New York, New York 10016, USA

In surface-enhanced Raman scattering experiments that use plasmonic nanostructures as substrates, the scattering spectrum contains a broad background usually associated with photoluminescence. This background exists above and below the frequency of the incident wave. The low-frequency part of this background is similar to the scattering spectrum of a plasmon nanoparticle, while the high-frequency part follows the Gibbs distribution. We develop a theory that explains experimentally observed features in both the high- and low-frequency parts of the photoluminescence spectrum from a unified point of view. We show that photoluminescence is attributed to the cascade Brillouin scattering of the incident wave by metal phonons under the plasmon resonance conditions. The theory is in good agreement with our measurements over the entire frequency range of the background.

## 1. INTRODUCTION

The term photoluminescence is commonly used for all types of light scattering for which spectrum is much broader than the spectrum of incident radiation. Photoluminescence of noble metals was first discovered back in 1969 [1] and was associated with a broad scattering spectrum resulting from incident argon laser light ranging from 457.9 to 514.5 nm and from Hg arc lamp light ranging from 300 to 400 nm. The first phenomenological theory of photoluminescence of noble metal nanostructures was put forward in Ref. [2] and subsequently developed in Refs. [3-5]. In this theory, it is assumed that photoluminescence is associated with the radiative *interband* recombination of sp-electrons and d-holes. Before recombining with a d-hole, an sp-electron may lose energy due to *intraband* transitions. This assumption means that photoluminescence should be observed at frequencies that are smaller than the frequency of incident light.

Interest in photoluminescence has been boosted by the surface-enhanced Raman scattering (SERS) measurements. In SERS, the analyte is on a noble metal substrate. The metal substrates can be made as an assembly of individual particles or as granular or corrugated surfaces. Special studies of such structures [6,7] show that the reflection spectrum of both a single particle [8,9] and a substrate made of granular plasmonic nanostructures [1,2] have a broad frequency-band background. Even though the background is observed both below and above the frequency of the incident wave, it is also referred to as photoluminescence. The integrated intensity of nanostructure photoluminescence (NSPL) strongly depends on the geometry of the plasmon nanostructure. For smooth metal surfaces, the NSPL intensity is extremely small. It is of the order of $10^{-10}$ smaller than the intensity of incident radiation. For granular substrates consisting of subwavelength spherical particles, this ratio is about $10^{-6}$, and for ellipsoid particles, it is of the order of $10^{-4}$ [2]. Since the Raman signal is very weak, NSPL may affect the results of Raman scattering measurements. The need to separate these two effects has inspired the studies of photoluminescence of nanostructures made of noble metals [8-10].

A detailed experimental study [2,6,7,9] has shown that NSPL of noble metal structures differs from the usual photoluminescence considered in Ref. [1]. First, in contrast to conventional luminescence, NSPL is even observed when the frequency of incident light, $\omega_0$, is lower than the frequencies of interband electron transitions [2,9] which, according to Ref. [1], are responsible for photoluminescence in metals. Second, the shapes of the NSPL spectra, $S_{\text{NSPL}}(\omega)$, are different above, $\omega > \omega_0$, and below $\omega < \omega_0$, the frequency of the incident wave. Specifically, above $\omega_0$, the NSPL intensity decreases with an increasing frequency $\omega$ according to the Gibbs distribution $\exp(-\hbar(\omega - \omega_0)/kT)$ [9]. Moreover, in the high-frequency part spectrum, the NSPL intensity increases with temperature [9]. Below $\omega_0$, $S_{\text{NSPL}}(\omega)$ is close to the *scattering spectrum for a plasmon nanostructure* (SSPNS), $S_p(\omega)$. This dependence arises when a plasmon structure is illuminated by a wave having a continuous white spectrum. The latter case was realized in Ref. [6], where plasmon nanoparticles were illuminated by short pulses.

It is natural to connect NSPL with the plasmon resonance of the roughness of a "smooth" surface of a nanostructure [6,7,11-13]. The elastic scattering of the incident wave on a nanostructure is usual Rayleigh scattering, which spectrum should coincide with the spectrum of incident radiation. However, NSPL excited by laser light has a broad spectrum whose shape is close to the SSPNS of the nanostructure. In particular, the NSPL spectrum has a maximum at the frequency close to the plasmonic resonance frequency. Thus, NSPL cannot be reduced to elastic scattering [9].

In Refs. [6,9], as an inelastic process, the *intraband* transitions of electrons inside the sp-band is considered. Since *intraband* dipole transitions are forbidden, the theory of Ref. [9], similar to Raman scattering, introduces a virtual level, to which an electron can be excited by capturing a photon, and then returning back. The process is considered as a creation of an electro-hole pair. To explain the Gibbs dependence of the high-frequency part of the spectrum, the following process has been employed. An electron thermally excited above the Fermi level absorbs the incident photon and gets excited to a virtual level. Then, this electron emits a plasmon and falls in an empty state slightly below the Fermi level. Since the states in the s-band are not all filled, in contrast to the interband transitions, where the hole energy is fixed, the electron has the opportunity to relax into a state with higher or lower energy, violating the elasticity of the process. The similarity of NSPL and the SSPNS has been explained by the fact that an excited electron-hole pair recombines emitting a plasmon. This explanation is in a qualitative agreement with experiment. Also, the Gibbs distribution of the number of electrons above the Fermi level explains the temperature dependence of the NSPL intensity in the high-frequency spectrum.

Although the consideration, based on the *intraband* transitions in the sp-band, qualitatively describes the plasmon resonant shape, to describe the low-frequency part of the NSPL spectrum and temperature dependence, an existence of an artificial virtual level has to be assumed.

In this paper, we propose a mechanism for the NSPL spectrum formation. To describe experiments, we use the recently developed approach that explains the inelastic Raman scattering as the direct interaction of electronic-driven oscillations with molecular nuclear vibrations [14-16]. We extend this approach to Brillouin scattering. The spontaneous Brillouin scattering of the electromagnetic (EM) field caused by this oscillation results in the excitation of a new oscillation of the EM field inside the nanoparticle at a shifted frequency (the Stokes shift). The field at the shifted frequency also undergoes Brillouin scattering launching the cascade of Brillouin scattering processes of an EM field inside the nanoparticle. The frequency of each new oscillation is shifted by the Stokes shift. Such a cascade process continues as long as the frequency of a new oscillation still falls into the SSPNS. In the Stokes part of the spectrum, the developed theory reproduces the SSPNS. We also show that the Gibbs distribution of the high-frequency intensity is due to the anti-Stokes Brillouin scattering. Thus, the inelastic Brillouin scattering is the inelastic process, the necessity of which has been indicated in Ref. [9]. All these effects are explained within a unified approach, without separate considerations of high- and low-frequency components. The theory is in good agreement with our measurements of the same sample over the entire frequency range of the background.

## II. PHOTOLUMINESCENCE SPECTRA FROM AU STOCHASTIC NANOSTRUCTURED FILMS. EXPERIMENT

To compare the results of the developed theory with experiment, we have to obtain our own experimental data because, to the best of our knowledge, in the literature [7,9,10], the experimental measurements obtained for the same sample, concern either the high- or low-frequency parts of the spectrum. The measurements for the whole frequency range of the NSPL spectrum are presented in Ref. [9]. However, in this paper, the frequency of the incident EM field is exactly equal to the plasmon resonance frequency.

In this section, we provide the results of the measurements of photoluminescence of granular gold films. The results are obtained for the entire spectrum of NSPL for different relations between the frequency of the incident field and the plasmon resonance frequency for the same nanoparticle. We use randomly nanostructured Au film (MATO S, manufactured by AtoiID, http://atoid.com). The stochastic nanopattern of a 200-nm thick Au film is formed by magnetron sputtering of the metal to an ultra-short laser pulse ablated soda-lime-silica glass substrate. To demonstrate the random character of the surface structure, SEM-images of two different fragments of an Au film surface are presented in Fig. 1 with the length scale indications of 500 nm and 5 μm. Features from a few tens of nanometers to almost a micrometer in size are observed. AFM-pictures of the surface show that its average roughness can be estimated as 0.4 μm. The samples provide distinct surface plasmon resonance characteristics for various excitation wavelengths.

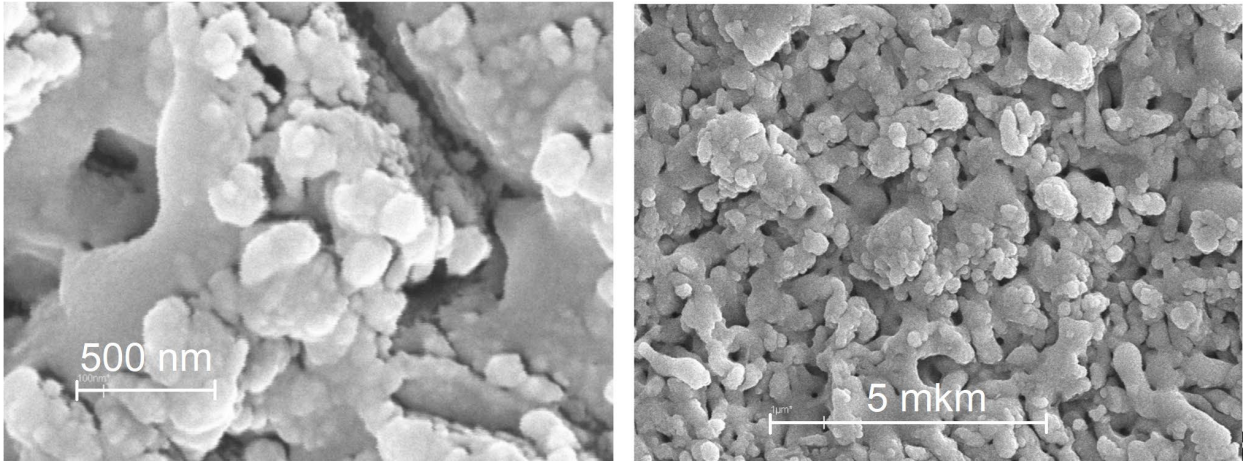

Fig. 1. SEM-images of the Au film surface area fragments.

To record photoluminescence spectra, we use a system described in Ref. [17]. The spectra presented below, are recorded employing a CW 10 mW 633 nm He-Ne laser (05-LHP-991, Melles-Griot), integrated into the system.

In Fig. 2, the thin blue lines represent photoluminescence spectra of the nanostructured Au surface which had been recorded in a ~1-μm diameter spatial point of the 12 μm × 12 μm fragment of the Au film surface at three different excitation intensities: 39, 75, and 150 μW/μm$^2$. One can clearly see the dip of ~600 cm$^{-1}$ width, provided by the notch filter near the excitation wavelength. Here, the spectral profiles are corrected for the detection system spectral efficiency, evaluated using the transmission curves of the optical elements, the grating diffraction efficiency, and the CCD quantum efficiency. These profiles represent the relative flux of the scattered radiation $I$ in photons/s. The position of the broad peak in the Stokes region varies within the range of about 500 cm$^{-1}$. The anti-Stokes part of the spectrum could be approximated as $I \sim \exp(-h(\omega - \omega_0)/k_B T)$, whith the temperature $T \approx 330 K$.

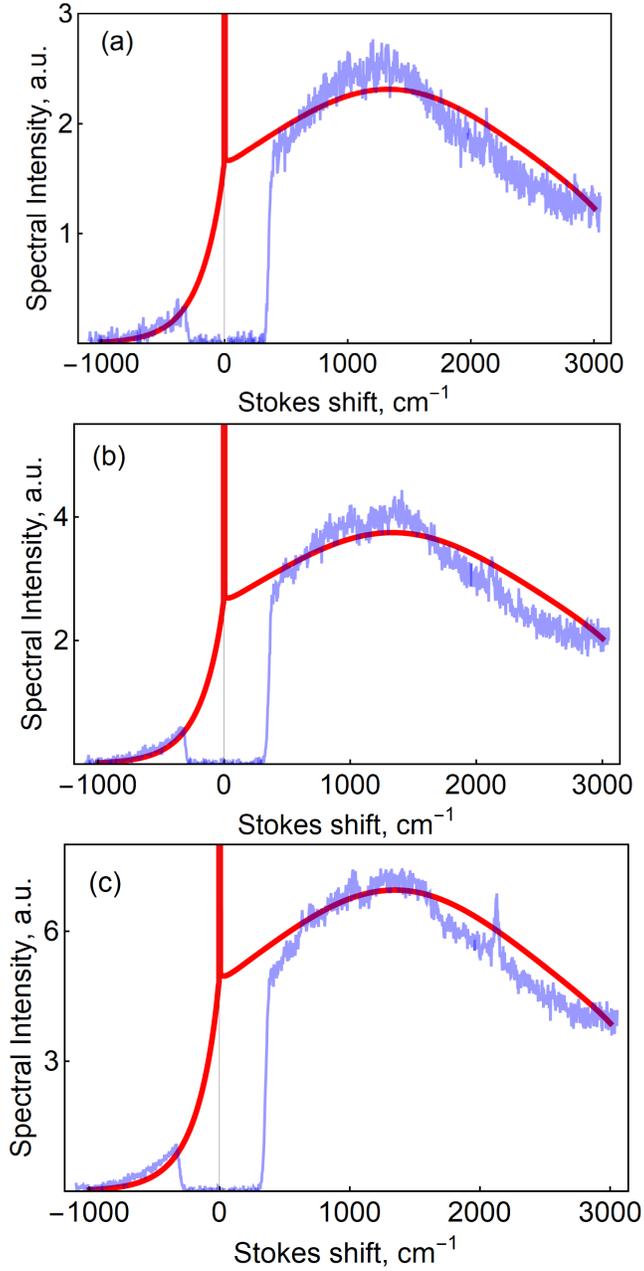

Fig. 2. The dependence of the intensity of scattered light (in arbitrary units) on the wavelength at pump rates of (a) $39\,\mu W$, (b) $75\,\mu W$, and (c) $150\,\mu W$. The experimental results are shown by the thin blue lines, the theoretical curves shown by the thick red lines are obtained by numerical calculations of Eqs. (2)–(4) and (19). Zero on the horizontal axis corresponds to the wavelength of incident light (633 nm).

## III. QUALITATIVE DESCRIPTION OF THE CASCADE BRILLOUIN PROCESS

As an inelastic mechanism leading to NSPL, we consider Brillouin scattering. A laser beam induces an EM field inside a plasmonic nanoparticle. We assume that the size of the nanoparticle is smaller than the skin depth. Consequently, the driven oscillations of the electric field inside the particle are uniform. Due to Brillouin scattering of this field, the EM field at the Stokes frequency appears. The frequency of this field is shifted by the phonon frequency. Because the latter is small in comparison to the width of SSPNS, the frequency of this excited field lies within the SSPNS.

The oscillations of the EM field caused by the incident wave and the scattered field cause the oscillations of electron density in metal and, consequently, the oscillations of the metal nuclei. Since the beat frequency of these fields coincides with the phonon frequency, the number of phonons resonantly increases, causing the increase in Brillouin scattering and in the intensity of the Stokes field. An increase in the number of quanta in the Stokes field, $n(\omega_{St})$, is determined by two processes. The first one is the stimulated excitation like that in the stimulated Raman scattering (SRS) [18], which is equal to $G_\downarrow n(\omega_{St}) n(\omega_0)$. Here $G_\downarrow$ denotes the rate of the increase in the intensity of the Stokes component (the symbol $\downarrow$ indicates that the process, in which the EM field with the lower frequency is excited by the field with the higher frequency - the Stokes process). The second process is the spontaneous Brillouin scattering, caused by vacuum zero-point vibrations with the Stokes frequency. This process is analogous to the excitation of the Stokes field in the Raman laser operating below the threshold. The rate of this process is $G_\downarrow n(\omega_0)$ [18]. Thus, the incident wave pumps the oscillation at the Stokes frequency, and the energy is accumulated in the EM field at the Stokes frequency. The dynamics of the Stocks intensity may be described by the rate equation

$$\dot{n}(\omega_{St}) = -\gamma_{rad}(\omega_{St}) n(\omega_{St}) + G_\downarrow n(\omega_0)(n(\omega_{St}) + 1), \qquad (1)$$

where $\gamma_{rad}(\omega_{St})$ is the rate of losses determining the width of the SSPNS. We can estimate the stationary value of $n(\omega_{St})$ as $n(\omega_{St}) \sim G_\downarrow n(\omega_0) / \gamma_{rad}(\omega_{St}) \ll 1$. Since $n(\omega_{St}) \ll 1$, the process of the spontaneous Brillouin scattering is much more intense than the stimulated emission.

The term $G_\downarrow n(\omega_{St}) n(\omega_0)$ can be expressed in the more standard form by using nonlinear susceptibility $\chi^{(3)}$. Namely, if instead of the number of quanta $n(\omega_{St})$, we consider the dynamics of the *intensity* of the field at the Stokes frequency, $I(\omega_{St})$, then the rate of the stimulated Brillouin scattering has the form $\sim \mathrm{Im}\chi^{(3)} I(\omega_{St}) I(\omega_0)$. The constants $G_\downarrow$ and $\chi^{(3)}$ are related by $G_\downarrow \approx \mathrm{Im}\chi^{(3)} \hbar \omega^2 / V$.

As we see, the spontaneous Brillouin scattering gives the main contribution to the process of photoluminescence. For this reason, we use the notation $G_\downarrow$ instead of $\chi^{(3)}$ to avoid the misleading association of photoluminescence with the stimulated process.

We assume that NSPL occurs due to a cascade of driven oscillations of the EM field inside the nanoparticle at different frequencies. Thus, we consider the nanoparticle as a resonator having the same line function shape as the SSPNS. Therefore, the incident field causes a driven oscillation of the field inside the nanoparticle with the frequency of the incident wave, $\omega_0$, and is accompanied by the oscillation of the dipole moment of the nanoparticle. That leads to Rayleigh scattering of the incident wave.

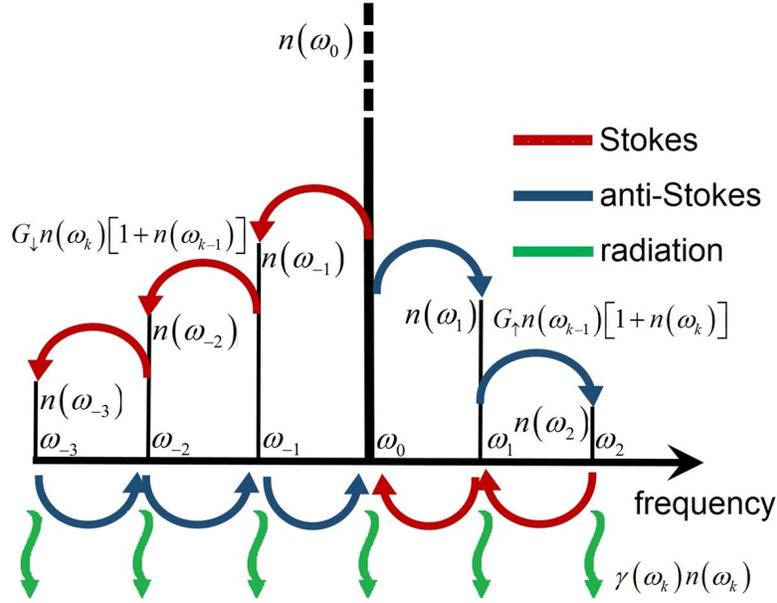

Fig. 3. The schematic of the interaction between the driven oscillations of the electric filed at various frequencies inside the nanoparticle. The $k$th driven oscillation dissipates with the rate $\sim \gamma(\omega_k) n(\omega_k)$ (wavy green lines) and exchanges the energy with the nearby driven oscillation at higher (blue lines) at lower (red lines) frequencies. The rates of corresponding exchanges of the energy are $G_\uparrow n(\omega_k)\left[1+n(\omega_{k+1})\right]$ and $G_\downarrow n(\omega_k)\left[1+n(\omega_{k-1})\right]$, respectively.

In Brillouin scattering, the Stokes shift is usually much smaller than the plasmon resonance linewidth. A *single* process, therefore, cannot explain the shape of the NSPL spectrum. To understand where the broad shape of NSPL comes from, one should take into account that the stationary Stokes field excites the next oscillations at the frequencies displaced by the same Stokes shift, $\omega_{ph}$, with respect to the Stokes frequency $\omega_0 - \omega_{ph}$. In turn, the EM field at the frequency $\omega_0 - \omega_{ph}$ excites the field at the frequency $\omega_0 - 2\omega_{ph}$, which then excites the field at the frequency $\omega_0 - 3\omega_{ph}$, and so on. These processes form a cascade of oscillations at frequencies inside the SSPNS (indicated by red arrows in Fig. 3, in which the cascade process is shown schematically). As for $n(\omega_{St})$, the process of the spontaneous Brillouin excitation for $n$ at shifted frequencies is dominant compared with stimulated excitations.

We enumerate oscillations with the subscript $k$. The frequencies of single anti-Stokes and Stokes shifts are $\omega_1 \equiv \omega_{aSt} = \omega_0 + \omega_{ph}$ and $\omega_{-1} \equiv \omega_{St} = \omega_0 - \omega_{ph}$, respectively, the frequencies of double anti-Stokes and Stokes shifts are $\omega_2 \equiv \omega_{aSt} + \omega_{ph} = \omega_0 + 2\omega_{ph}$ and $\omega_{-2} \equiv \omega_{St} - \omega_{ph} = \omega_0 - 2\omega_{ph}$, respectively, and so on. The number of quanta in the $k$th oscillation is denoted by $n(\omega_k)$.

The radiated intensity is proportional to the product of the squared dipole moment and the number of quanta. The dependence of the dipole susceptibility of the nanoparticle on the frequency has the Lorentzian shape, see Fig. 4 (this figure is obtained in Sec. V). Therefore, the amplitude of the emission spectrum (Fig. 4, red circles) differs from the amplitude of the number of oscillation quanta (Fig. 4, blue circles) by the factor, which has a maximum at the plasmon resonance frequency (Fig. 4, green line). Taking into account the decay of oscillations and the dispersion factor with a resonant frequency dependence, we obtain the radiation spectrum with the maximum shifted toward the incident radiation frequency (the difference between the red circles and the green line in Fig. 4. This shift is observed in experiment [8].

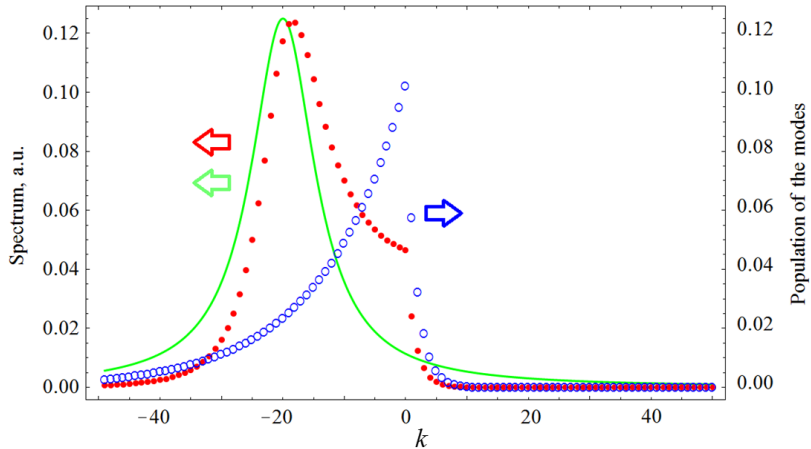

Fig. 4. The frequency dependencies of the quasiparticle population of the mode $n_n^{st}$ (blue circles) and the quasiparticle radiation rate $\gamma_{rad}(\omega_k)$ (the solid green line) on the mode number. Red circles represent the dependence of the spectrum $S(\omega_n)$, defined by Eq. (22), on the frequencies $\omega_n$. System parameters are $\omega_0 = 633$nm, $T = 300$K, $\gamma/\omega_0 = 0.01$, $G_\downarrow/\omega_0 = 0.3$, $G_\uparrow = G_\downarrow \exp(-\Delta\omega/T) = 0.3 G_\downarrow$, and $P/\omega_0 = 0.017$.

Along with the excitation of the EM field at the Stokes frequency, the field is also excited at the anti-Stokes frequency (blue arrows in Fig. 3). This results in the energy flow from the field at the frequency $\omega_k$ to the field at the frequency $\omega_k + \omega_{ph}$.

The qualitative description of the mechanism for NSPL of metals presented above does not take into account the cascade of anti-Stokes scatterings. Consequently, it does not describe the shape of the spectrum above the frequency of the incident EM field, which is a result of such a cascade. For its description, a consistent quantum mechanical approach is necessary. This description is developed in the next section.

## IV. QUANTUM MECHANICAL EQUATIONS DESCRIBING CASCADE BRILLOUIN PROCESS

As we discuss in the previous section, the driven oscillation excited by the external field with the frequency $\omega_0$ is amplified due to the closeness of $\omega_0$ to $\omega_{pl}$. Due to the nonlinear Brillouin scattering, the EM field of this driven oscillation at the frequency $\omega_0$ induces the next driven oscillation with the frequency $\omega_0 - \omega_{ph}$. We denote the complex amplitude of a driven oscillation at the frequency $\omega_k$ as $a(\omega_k, t)$.

Brillouin scattering also contributes to the fields at anti-Stokes frequencies. Thus, there as a process of the energy redistribution among the driven oscillations. To describe the dynamics of the cascade Brillouin energy transport in NSPL, we should take into account the quantum properties of the phenomena.

For this purpose, we use the procedure of macroscopic quantization [19,20] that includes quantization of the EM field and collective electron oscillations in the plasmon structure, as well as the modes of reservoirs with which these electrons interact. Such reservoirs can be phonons, impurities, etc. This theory relies only on the general linear properties of the medium (the linear permittivity) and does not require complicated first-principle calculations. The eigenmodes of such a system are collective oscillation states of electrons and reservoir modes. The eigenmodes (quasiparticles) can be found using the Fano diagonalization procedure [19-21]. Each quasiparticle has a dipole moment. The closer the frequency of a quasiparticle to the plasmon resonance frequency, the greater the dipole moment of this quasiparticle. As shown in Appendix A, the frequency dependence of the squared dipole moment of the quasiparticle coincides with the SSPNS. Thus, we obtain quasiparticles at each frequency that belongs to the SSPNS. Each quasiparticle with the frequency $\omega_k$ is described by the annihilation and creation operators, $\hat{a}(\omega_k, t)$ and $\hat{a}^\dagger(\omega_k, t)$; the operator of the number of quasiparticle is $\hat{n}(\omega_k, t) = \hat{a}^\dagger(\omega_k, t) \hat{a}(\omega_k, t)$. Thus, each quasiparticle is identified with an eigensolution corresponding to any driven oscillation considered above.

In Appendix B, we derive the equations of motion for the expected values of operators $\langle \hat{a}(\omega_0,t) \rangle \equiv a(\omega_0,t)$, $\langle \hat{a}(\omega_k,t) \rangle \equiv a(\omega_k,t)$ and $\langle \hat{n}(\omega_k,t) \rangle \equiv n(\omega_k,t)$. $a(\omega_0,t)$ describes the coherent driven oscillation caused by the external field, $a(\omega_k,t)$ where $k \neq 0$, describes incoherent NSPL, and $n(\omega_k,t) \equiv \langle \hat{n}(\omega_k,t) \rangle$ is the number of quasiparticles excited at the frequency $\omega_k$. The latter quantity up to the factor $\hbar\omega_k$ determines the total energy (coherent and incoherent) in the corresponding mode. The corresponding equations have the form:

$$\frac{da(\omega_0,t)}{dt} = \left(-i\omega_0 - \gamma_{\text{rad}}(\omega_0)/2\right)a(\omega_0,t) - i\Omega_{\text{ex}}\exp(-i\omega_0 t)$$
$$+ \frac{a(\omega_0,t)}{2}\sum_m \left\{ G_{\omega_m,\omega_0} n(\omega_m,t) - G_{\omega_0,\omega_m}\left[n(\omega_m,t)+1\right]\right\}, \quad (2)$$

$$\frac{da(\omega_k,t)}{dt} = \left(-i\omega_k - \gamma_{\text{rad}}(\omega_k)/2\right)a(\omega_k,t)$$
$$+ \frac{a(\omega_k,t)}{2}\sum_m \left\{ G_{\omega_m,\omega_k} n(\omega_m,t) - G_{\omega_k,\omega_m}\left[n(\omega_m,t)+1\right]\right\}, \quad k\neq 0, \quad (3)$$

$$\frac{dn(\omega_k,t)}{dt} = -\gamma_{\text{rad}}(\omega_k) n(\omega_k,t)$$
$$+ \sum_m \left\{ n(\omega_m,t)\left[n(\omega_k,t)+1\right]G_{\omega_m,\omega_k} - n(\omega_k,t)\left[n(\omega_m,t)+1\right]G_{\omega_k,\omega_m}\right\}, \quad k\neq 0, \quad (4)$$

where $\gamma_{\text{rad}}(\omega_k)$ is the rate of radiative losses of the quasiparticle with the eigenfrequency $\omega_k$, $\Omega_0 = -\mathbf{d}(\omega_0)\cdot\mathbf{E}_0/\hbar$ is the interaction constant between the external field with the amplitude $\mathbf{E}_0$ and the quasiparticle with the frequency $\omega_0$ and the dipole moment $\mathbf{d}(\omega_0)$. For the nanoparticle of the radius $R$, the dipole moment of the $k$th quasiparticle is determined as,

$$\mathbf{d}(\omega) = 4\pi R^3 \sqrt{\frac{\Delta\omega\hbar\varepsilon_0}{\pi R^3}} \frac{\sqrt{\text{Im}\,\varepsilon_{\text{M}}^{(lin)}(\omega)}}{\left|\hat{\varepsilon}_{\text{M}}^{(lin)}(\omega)+2\right|}\mathbf{e}_d,$$

where $\mathbf{e}_d$ is the unit vector in the direction of the dipole moment, and $\hat{\varepsilon}_{\text{M}}^{(lin)}(\omega) = \text{Re}\,\varepsilon_{\text{M}}^{(lin)}(\omega) + i\,\text{Im}\,\varepsilon_{\text{M}}^{(lin)}(\omega)$ is the linear part of metal permittivity.

Note that if in Eq. (4) we put $\omega_k = \omega_{\text{St}}$, then in the sum, only the term with $m=0$ remains; then, replacing $G_{\omega_0,\omega_{\text{St}}}$ by $G_\downarrow$, we obtain Eq. (1).

## V. ESTIMATION OF THE RATE OF BRILLOUIN SCATTERING

Before analyzing the system of Eqs. (2)–(4), we qualitatively evaluate the interaction constant, $G_{\omega_m,\omega_k}$, between the EM field modes due to the interaction with the phonon reservoir for various geometries. As mentioned in the introduction, the intensity of Brillouin scattering on a metal surface is smaller by order of magnitude compared with photoluminescence from a nanoparticle. The model developed in the previous section is quite general and can be applied to Brillouin scattering on a metal surface. The difference between the scattering at a surface and a nanoparticle is in the constants of the interaction between the field mode and the phonon mode. In this section, we qualitatively evaluate these interaction constants.

The interaction constant $w_{nml}$ between the $n$th and $m$th field modes and the $l$th phonon is derived in Appendix A; it has the form

$$w_{nml} = \tilde{w}_l \int_V d^3\mathbf{r}\, \Lambda_n^{(0)*}(\mathbf{r}) \Lambda_m^{(0)}(\mathbf{r}) C_l^*(\mathbf{r}), \tag{5}$$

where $\Lambda_m^{(0)}(\mathbf{r})$ is the electric field distribution of the eigenmode of the EM field and $C_l^*(\mathbf{r})$ is the distribution function of the eigendisplacement of phonons. The integration in Eq. (5) is carried out over the volume of the medium. In the case of a nanoparticle, inside the medium, $\Lambda_m^{(0)}(\mathbf{r}) \sim \sqrt{\hbar \omega_m / V}\, \mathbf{e}_d$ ($\mathbf{e}_d$ is the unit vector in the direction of the EM field), $C_l(\mathbf{r}) \sim \sqrt{\hbar / M \omega_l V}\, \exp(i\mathbf{k}_l \cdot \mathbf{r})$, where $\mathbf{k}_l$ is the wavevector of $l$th phonon. The substitution of $C_l^*(\mathbf{r})$ into the Eq. (5) gives

$$w_{nml} \sim w \sqrt{\frac{\hbar^3 \omega_n \omega_m}{MV \omega_l^{ph}}} \frac{1}{V k_l^3} \left( \sin(k_l R) - k R \cos(k_l R) \right), \tag{6}$$

The constant $G_{\omega_n, \omega_m}$, which determines the energy flow from the $m$th EM field mode to the $n$th mode, is connected with the interaction constant through the relation [see Appendix B, Eq. (B6)]:

$$G_{\omega_m, \omega_n} \approx \pi \sum_l |w_{nml}|^2 \left(1 + N(\omega_l^{ph})\right) \delta(\omega_n - \omega_m - \omega_l^{ph})$$

$$= \frac{\pi V}{(2\pi)^3} \int d^3\mathbf{k}\, |w_{nm}(\mathbf{k})|^2 \left(1 + N(\omega_\mathbf{k}^{ph})\right) \delta(\omega_n - \omega_m - \omega_\mathbf{k}^{ph})$$

$$\approx \frac{w^2 \hbar^3 \omega_n \omega_m V}{2\pi M V c_{ph}^3} \int d\omega^{ph} \frac{\omega^{ph}}{(Vk_l^3)^2} \left( \sin(k_l R) - k R \cos(k_l R) \right)^2 \left(1 + N(\omega^{ph})\right) \delta(\omega_n - \omega_m - \omega^{ph})$$

$$= \frac{w^2 \hbar^3 \omega_n \omega_m}{2\pi M c_{ph}^3} (\omega_n - \omega_m)\left(1 + N(\omega_n - \omega_m)\right) \tag{7}$$

$$\times \left( \sin\left(\frac{(\omega_n - \omega_m) R}{v_s}\right) - \frac{(\omega_n - \omega_m) R}{v_s} \cos\left(\frac{(\omega_n - \omega_m) R}{v_s}\right) \right)^2 (Vk_l^3)^{-2}$$

$$= G_0 (\omega_n - \omega_m)\left(1 + N(\omega_n - \omega_m)\right)$$

$$\times \left( \sin\left(\frac{(\omega_n - \omega_m) R}{v_s}\right) - \frac{(\omega_n - \omega_m) R}{v_s} \cos\left(\frac{(\omega_n - \omega_m) R}{v_s}\right) \right)^2 (Vk_l^3)^{-2}$$

where we denote $w^2 = |w_{nm}(\omega_n - \omega_m)|^2$ and $G_0 = (w^2 \hbar^3 \omega_n \omega_m)/(2\pi M c_{ph}^3)$. The dependence of $G_{\omega_n,\omega_m}$ on the frequency difference $\Delta\omega \equiv \omega_n - \omega_m$ is shown in Fig. 5.

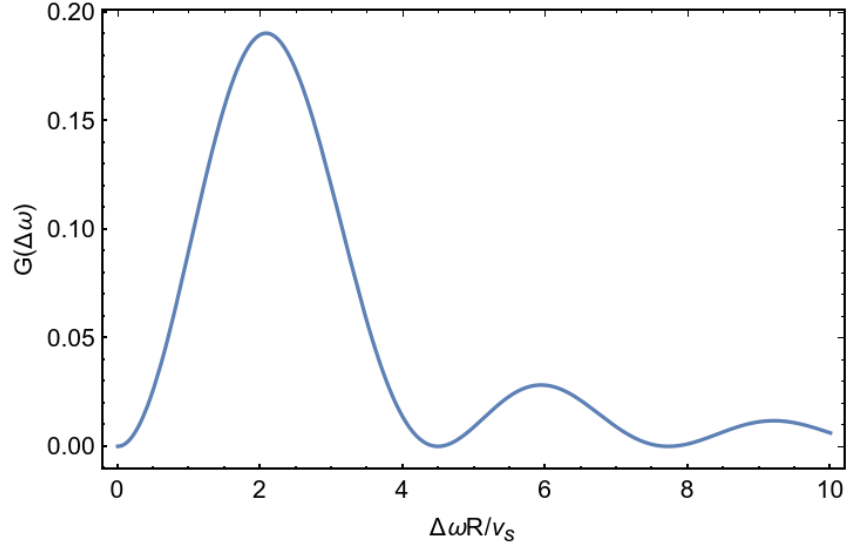

Fig. 5. The dependence of $G_{\omega_n,\omega_m}$ on the frequency difference $\Delta\omega \equiv \omega_n - \omega_m$

The first maximum of $G_{\omega_n,\omega_m}$ corresponds to the frequency difference $\Delta\omega \sim v_s/R$ and the wavevector $k_l \sim 1/R$, i.e., this maximum corresponds to the bulk resonance at the metal sphere [11]. Thus, if one considers $G_{\omega_n,\omega_m}$ for the frequency difference $\omega_n - \omega_m \sim v_s/R$, the factor $V k_l^3$ is of the order of unity, and $G_{\omega_n,\omega_m}$ does not depend on the volume.

The situation is different in the case of the metal surface. The EM field penetrates the metal to the skin depth $\delta$. We estimate the field distribution inside the metal as $\Lambda_m^{(0)}(\mathbf{r}) \sim \mathbf{e}\sqrt{\hbar\omega_m/V} \exp(i\mathbf{k}_\parallel \cdot \mathbf{r}_\parallel)\exp(-z/\delta)$ ($z$ is the direction from the metal surface into the nanoparticle). Substituting this distributions into Eq. (5) we obtain

$$w_{nml} \sim w \sqrt{\frac{\hbar^3 \omega_n \omega_m}{MV\omega_l^{ph}}} \frac{\delta}{V(1-ik_z^{ph}\delta)} \int_A d^2\mathbf{r} \exp\left(i\left(\mathbf{k}_\parallel^n - \mathbf{k}_\parallel^m - \mathbf{k}_\parallel^{ph}\right)\cdot \mathbf{r}_\parallel\right), \qquad (8)$$

where $A$ is the surface area. The substitution of Eq. (8) into the expression for $G_{\omega_n,\omega_m}$ gives

$$G_{\omega_m,\omega_n} \approx \pi \sum_l |w_{nml}|^2 \left(1+N(\omega_l^{ph})\right)\delta(\omega_n - \omega_m - \omega_l^{ph})$$

$$= \frac{\pi V}{(2\pi)^3} \int d^3\mathbf{k}^{ph} \left|w_{nm}(\mathbf{k}^{ph})\right|^2 \left(1+N(\omega_\mathbf{k}^{ph})\right)\delta(\omega_n - \omega_m - \omega_\mathbf{k}^{ph}) \qquad (9)$$

$$\approx \frac{w^2 \hbar^3 \omega_n \omega_m}{8\pi^2 M}\left(\frac{\delta}{L}\right)^2 \frac{1}{V}\int d^3\mathbf{k}^{ph} \frac{\left(1+N(\omega^{ph})\right)\delta\left(\mathbf{k}_\parallel^n - \mathbf{k}_\parallel^m - \mathbf{k}_\parallel^{ph}\right)\delta(\omega_n - \omega_m - \omega^{ph})}{\omega^{ph}\left(1+(k_z^{ph}\delta)^2\right)},$$

where $L$ is the thickness of the medium. The $\delta$-function inside the integral reflects the fact that the EM field modes with the frequencies $\omega_n$ and $\omega_m$ and the parallel wavevector components $\mathbf{k}_\parallel^n$ and $\mathbf{k}_\parallel^m$ interact only with the phonon which frequency equals $\omega^{ph} = \omega_n - \omega_m$ and parallel wavevector component equals $\mathbf{k}_\parallel^{ph} = \mathbf{k}_\parallel^n - \mathbf{k}_\parallel^m$. It can be shown [22,23] that for the given incident and reflected waves of the EM field, only zero or one phonon can satisfy such conditions. Denoting the frequency of the corresponding phonon as $\tilde{\omega}^{ph}$ we can transform Eq. (9) into

$$G_{\omega_m,\omega_n} \sim \frac{w^2 \hbar^3 \omega_n \omega_m}{8\pi^2 M} \left(\frac{\delta}{L}\right)^2 \frac{1}{V} \frac{\left(1 + N(\tilde{\omega}^{ph})\right)}{\tilde{\omega}^{ph}\left(1 + \left(\tilde{k}_z^{ph}\delta\right)^2\right)}. \tag{10}$$

Equation (10) describes the rate of the energy transfer from the wave with the wavevector $\mathbf{k}_n$ to the wave with the wavevector $\mathbf{k}_m$. This value is inversely proportional to the volume $V$. If we sum up the signal of Brillouin scattering over the wavevector interval $\Delta \mathbf{k}_m$ and frequency interval $\Delta \omega_m$, this volume is reduced, and we obtain

$$G_{\Delta\omega} = \sum_{\Delta \mathbf{k}_m} G_{\omega_m,\omega_n} = \frac{V}{(2\pi)^3} \int_{\Delta \mathbf{k}_m} d^3 \mathbf{k}_m G_{\omega_m,\omega_n} \sim \frac{w^2 \hbar^3 \omega_n \omega_m}{64\pi^5 M} \left(\frac{\delta}{L}\right)^2 \frac{\omega_m \Delta \omega_m \left(1 + N(\tilde{\omega}^{ph})\right)}{c^3 \tilde{\omega}^{ph}\left(1 + \left(\tilde{k}_z^{ph}\delta\right)^2\right)}. \tag{11}$$

Compared with the case of nanoparticle, the constant $G_{\Delta\omega}$ contains the factor $(\delta/L)^2$. Because the phonons have a finite coherence length, one should replace the total thickness of the medium $L$ by the phonon coherence length $L_{coh}$. As a result, the rate of Brillouin scattering at the metal surface is smaller than that in the metal nanoparticle by the factor $(\delta/L_{coh})^2$. This factor for the metal at room temperature is $(\delta/L_{coh})^2 \sim 10^{-3}$, which explains the enhancement of Brillouin scattering in the nanoparticle.

## VI. PHOTOLUMINESCENCE SPECTRUM

Equation (3) has a trivial stationary solution, $a(\omega_k,t) = 0, k \neq 0$. This solution indicates that the quasiparticles at the frequencies $\omega_k \neq \omega_0$ are incoherent. $a(\omega_k,t) \sim a^{st}(\omega_k)\exp(-i\omega_k t)$ with $a(\omega_k) \neq 0, k \neq 0$, which is realized when the condition $\sum_m \{G_{\omega_m,\omega_k} n(\omega_m,t) - G_{\omega_k,\omega_m}(n(\omega_m,t)+1)\} = \gamma_{rad}(\omega_k)$ is fulfilled. This solution requires that $n(\omega_k,t) \neq 0$. This corresponds to the beginning of self-oscillations, which takes place in the Raman laser with high-$Q$ resonators [24,25]. In the case when a plasmonic nanoparticle plays the role of the resonator, the beginning of self-oscillations can be achieved only at high-power pulse pumping [26]. Here we do not consider such a situation and restrict ourselves to consideration of the case $a(\omega_k,t)=0$, $k \neq 0$, which is relevant to our experiment.

The values $G_{\omega_m,\omega_k}$ are the rates of scattering of the quasiparticle from the state with the frequency $\omega_k$ to the state with the frequency $\omega_m$. The term $\sim G_{\omega_m,\omega_k} n(\omega_m,t) n(\omega_k,t)$ describes the process of the stimulated Brillouin excitation (like in SRS), while $\sim G_{\omega_m,\omega_k} n(\omega_m,t)$ describes the process of spontaneous excitation. It can also be shown that $G_{\omega_m,\omega_k} \sim \mathrm{Im}\,\chi^{(3)}(\omega_m - \omega_k)$.

Since the energy transfer coefficient, $G_{\omega_m,\omega_k}$, is rather small, we assume that $n(\omega_k) \ll n(\omega_0)$ for any $k \neq 0$. For a stationary process, when $t \gg t_{st}$, we can neglect time derivatives in the left-hand side of Eqs. (2)-(4). Using this assumption, we find the equation for the stationary values of $a^{st}(\omega_0,t)$ and $n(\omega_k)$:

$$a^{st}(\omega_0,t) = \frac{-i\Omega_{ex} \exp(-i\omega_0 t)}{\gamma_{rad}(\omega_0) + \sum_m G_{\omega_m,\omega_0}}, \qquad (12)$$

$$-\gamma_{rad}(\omega_k) n(\omega_k,t_{st})$$
$$+ \sum_m \left\{ G_{\omega_k,\omega_m} n(\omega_m,t_{st})[1+n(\omega_k,t_{st})] - G_{\omega_k,\omega_m} n(\omega_k,t_{st})[1+n(\omega_m,t_{st})] \right\} = 0, \ k \neq 0. \qquad (13)$$

Now we are ready to calculate the whole emission spectrum, $S(\omega)$, which includes both NSPL spectrum, $S_{NSPL}(\omega)$, and Rayleigh scattering, $S_R(\omega)$, of a plasmon nanoparticle. Using the standard approach [27,28] we obtain:

$$S(\omega) = \frac{1}{\pi} \sum_k \hbar \omega_k \gamma_{rad}(\omega_k) \mathrm{Re} \int_0^{+\infty} d\tau \langle \hat{a}^\dagger(\omega_k,t_{st}) \hat{a}(\omega_k,t_{st}+\tau) \rangle \exp(i\omega\tau). \qquad (14)$$

To complete the calculations, we need to find two-time correlators $\langle \hat{a}^\dagger(\omega_0,t)\hat{a}(\omega_0,t+\tau) \rangle$ and $\langle \hat{a}^\dagger(\omega_k,t_{st})\hat{a}(\omega_k,t_{st}+\tau) \rangle$. Note that the latter correlator is not equal to zero even though $\langle \hat{a}(\omega_k,t_{st}+\tau) \rangle = 0$. To do this, we use the quantum regression theorem [28], which states that two-time correlators of an operator are described by the same equations as the one-time correlator. In our case, these correlators are governed by Eqs. (2) and (3), respectively. Assuming that $n(\omega_k,t) \ll 1$ and $\gamma_{rad}(\omega_0) \gg G_{\omega_k,\omega_m}$ we obtain:

$$\frac{d}{d\tau} \langle \hat{a}^\dagger(\omega_0,t)\hat{a}(\omega_0,t+\tau) \rangle = (-i\omega_0 - \gamma_{rad}(\omega_0)) \langle \hat{a}^\dagger(\omega_0,t)\hat{a}(\omega_0,t+\tau) \rangle$$
$$- i |a(\omega_0,t_{st})|^2 \Omega_{ex} \exp(-i\omega_0\tau), \qquad (15)$$

$$\frac{d}{d\tau} \langle \hat{a}^\dagger(\omega_k,t)\hat{a}(\omega_k,t+\tau) \rangle = (-i\omega_k - \gamma_{rad}(\omega_k)) \langle \hat{a}^\dagger(\omega_k,t)\hat{a}(\omega_k,t+\tau) \rangle, \ k \neq 0. \qquad (16)$$

The solution to Eqs. (15) and (16) for $t \to +\infty$ are

$$\langle \hat{a}^\dagger(\omega_0, t_{st}) \hat{a}(\omega_0, t_{st} + \tau) \rangle = |a(\omega_0, t_{st})|^2 \exp(-i\omega_0 \tau), \qquad (17)$$

$$\langle \hat{a}^\dagger(\omega_k, t_{st}) \hat{a}(\omega_k, t_{st} + \tau) \rangle = n(\omega_k, t_{st}) \exp\left[(-i\omega_k - \gamma_{rad}(\omega_k))\tau\right], \ k \neq 0, \qquad (18)$$

where $a(\omega_0, t_{st})$ is determined by Eq. (12), and the quantity $n(\omega_k, t_{st})$, could be found from Eq. (13).

Equation (7) contains both contributions $S_R(\omega)$ of Rayleigh scattering [see Eq. (17)] and of photoluminescence $S_{NSPL}(\omega)$ [see Eq. (18)]. Combining Eqs. (14) and (18) one can find the NSPL spectrum $S_{NSPL}(\omega)$:

$$S_{NSPL}(\omega) = \frac{1}{\pi} \sum_{k \neq 0} \frac{\hbar \omega_k \gamma_{rad}(\omega_k)}{\gamma_{rad}^2(\omega_k) + (\omega_k - \omega)^2} n(\omega_k, t_{st}). \qquad (19)$$

In Eq. (19), the multiplier $\hbar \omega_k \gamma_{rad}(\omega_k)$ indicates how efficiently the quasiparticle shines at the frequency $\omega_k$. As noted above, $\gamma_{rad}(\omega_k) \propto 1/|\varepsilon_M(\omega_k) + 2|^2$ that is, the modes whose frequencies are located near the plasmon resonance frequency $\omega_{pl}$ shine most efficiently.

## V. TOY MODEL

To illustrate the developed approach, we apply the results obtained in the previous section to a simple model, the solution of which can be obtained analytically. We consider modes whose eigenfrequencies $\omega_k$ are equidistant ($\omega_{k+1} - \omega_k = \Delta \omega = \text{const}$); we also assume that the energy transfer can occur directly between neighboring modes only. The stationary solution can be found by solving Eq. (13). For simplicity, we assume that $\gamma_{rad}(\omega_k) = \gamma$ for all modes, $n_{\omega_k}^{st} \ll 1$, and the external field excites only the mode with the index 0. In this case, the equation for the stationary distribution of the mode population (13) takes the form:

$$-\gamma n_{st}(\omega_k) + G_\downarrow n_{st}(\omega_{k+1}) + G_\uparrow n_{st}(\omega_{k-1}) - G_\downarrow n_{st}(\omega_k) - G_\uparrow n_{st}(\omega_k) + P\delta_{\omega_0, \omega_k} = 0, \qquad (20)$$

where we designate $G_\downarrow = G_{\omega_{k-1}, \omega_k}$ and $G_\uparrow = G_{\omega_{k+1}, \omega_k}$ for all modes. The rates $G_\downarrow$ and $G_\uparrow$ satisfy Kubo-Martin-Schwinger relation $G_\uparrow / G_\downarrow = \exp(-\Delta \omega / T)$. The quantity $P$ determines the intensity of the external field acting on the $\omega_0$-mode. The solution of Eq. (20) has the form:

$$n_{st}(\omega_k) = \begin{cases} n_{st}(\omega_{ex}) e^{(\omega_k - \omega_0)\lambda_-}, & k > 0, \\ n_{st}(\omega_{ex}) e^{(\omega_k - \omega_0)\lambda_+}, & k < 0, \end{cases} \qquad (21)$$

where

$$\lambda_\pm = \frac{1}{\Delta \omega} \ln\left[\left(\gamma + G_\downarrow + G_\uparrow \pm \sqrt{(\gamma + G_\downarrow + G_\uparrow)^2 - 4 G_\downarrow G_\uparrow}\right) \Big/ 2 G_\downarrow\right],$$

and the population $n_{st}(\omega_0)$ is the number of quasiparticles that are excited at the frequency of the external field and lead to Rayleigh scattering [see Eq. (12)]. Note that when $G_\downarrow \gg \gamma$, $\lambda_- \sim \frac{1}{\Delta\omega}\ln[G_\uparrow/G_\downarrow] \sim -1/T$ and the number of quanta at the high-frequency part decreases according to Gibbs distribution, $n_{st}(\omega_k) \sim \exp(-(\omega_k - \omega_0)/T)$.

The assumption that $\gamma_{rad}(\omega_k)$ is frequency independent is made for the clarity of calculations. In the toy model, to obtain the radiation spectrum, we need to use more plausible Lorentzian line shape

$$\gamma_{rad}(\omega_k) = \frac{\gamma_{pl}^2/\pi}{(\omega - \omega_{pl})^2 + \gamma_{pl}^2}\Delta\omega,$$

where $\gamma_{pl}$ is the experimental value of the SSPNS width (see Appendix A). The energy emitted by the system, $S(\omega_n)$, at the frequency $\omega_n$ is given by the expression:

$$S(\omega_k) = \gamma_{rad}(\omega_k)n_{st}(\omega_k). \tag{22}$$

The spectrum obtained by using Eq. (22), is shown in Fig. 4 by the red circles. We see that the spectrum has a maximum in the neighborhood of the plasmon resonance frequency in the low-frequency range. In the high-frequency range, the intensity of the spectrum decreases according to the Gibbs distribution because $S(\omega_k) = \gamma_{rad}(\omega_k)n_{st}(\omega_k) \sim \exp(-(\omega_k - \omega_0)/T)$.

## VII. RADIATION SPECTRUM IN EXACT MODEL

To find the spectrum of NSPL, we solve the complete system of equations using computer simulation of Eqs. (2)–(4). In numerical modeling, a solution to Eqs. (2)–(4) is found for a discrete finite set of frequencies $\{\omega_k\}$. This set is defined according to the formula:

$$\omega_k = \omega_{min} + \frac{k-1}{N-1}(\omega_{max} - \omega_{min}), \quad k = 1,...,N, \tag{23}$$

where to compare to the experiment, we consider the frequency region that covers the plasmon resonance line with the minimum and maximum frequencies $\omega_{min} = 1.5\,\text{eV}$ and $\omega_{max} = 2.4\,\text{eV}$, 12100 cm-1 respectively, and the total number of frequencies is $N = 400$. The frequency of the plasmon resonance is taken from the experimental data shown in Section 2 (Fig. 2, the blue line), $\omega_{pl} = 1.75\,\text{eV}$. It is assumed that the external EM field has the frequency $\omega_0 = 1.96\,\text{eV}$, and the radius of the nanoparticle $R = 50\,\text{nm}$. Once $n_k^{st}$ are found, we substitute $n_k^{st}$ into Eq. (19) and define the spectrum of NSPL. The results of computer simulation of Eqs. (2)–(4) give the mode occupation numbers $n_k^{st}$. To determine the NSPL spectrum, $S_{pl}(\omega)$, these numbers are substituted into Eq. (19) . Note that in noble metals, the phonon frequencies form the allowed band lie inside the region between 0 and $\omega_{ph}^{max}$, which is about several THz [29,30], therefore, driven oscillations inside the interval $\left(\omega_{ex}-\omega_{ph}^{max},\omega_{ex}+\omega_{ph}^{max}\right)$ may be excited. Consequently, instead of a discrete spectrum, we obtain a continuous one, shown in Fig. 2 by the red line.

This spectrum reproduces the main characteristics observed in the experiment: the shape of the plasmon resonance in the low-frequency region and the Gibbs distribution in the high-frequency region. It also reproduces the shape of the SSPNS in the low-frequency region and the Gibbs distribution in the high-frequency region. Note that a direct consequence of Eqs. (2)–(4) is an increase in the intensity of the high-frequency part of the photoluminescence spectra with increasing temperature (see Fig. 6). Such an increase has been observed in experiment [9].

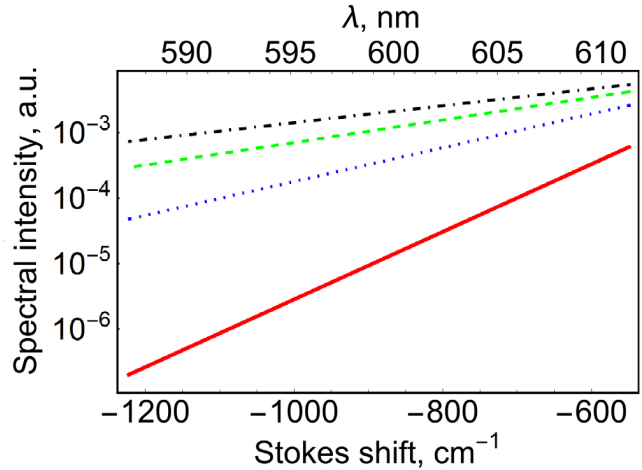

Fig. 6. The high-frequency part of the luminescence spectrum at temperatures 100K (the solid red line), 200K (the dotted blue line), 300K (the dashed green line), and 400K (the dot-dashed black line). For all temperatures, $S(\omega)$ falls off exponentially in agreement with the Gibbs distribution.

Let us compare experimental and theoretical results. In numerical simulation, we use the constant $G_0$ from Eq. (7) as a fitting parameter. The best fit shown in Fig. 2 is obtained for $G_0=10^{-8}s^{-1}$. One can see that our model correctly describes both the decrease of the amplitude of the high-frequency part of NSPL spectra that changes according to the Gibbs distribution, $\sim\exp\left(-h(\omega-\omega_0)/k_BT\right)$, and the maximum in the low-frequency part of NSPL spectra at the frequency of the plasmon resonance. Note that in experiment, the Rayleigh peak in the lower part of the spectrum is usually cut out by a filter, and only the background is visible.

## VIII. CONCLUSIONS

The proposed mechanism for photoluminescence of nanostructures made of noble metals correctly describes the spectrum for the entire range of frequencies. NSPL is attributed to the cascade Brillouin scattering of the incident wave by metal phonons under the plasmon resonance conditions. The mechanism consists of the cascade Brillouin scattering of the driven oscillation of the electric field inside the SSPNS. The energy transport occurs from the modes with high frequencies toward the modes with low frequencies. The elementary process in the cascade is the one in which an excitation at one frequency gives rise to a cascade of excitations due to the interaction of the electron density with phonons of the metal. The cascade Brillouin scattering leads to the broadening of the NSPL spectrum far beyond the maximum phonon frequency in metal because the energy transfer between driven oscillations occurs sequentially.

The developed theory is in good agreement with the experimentally observed NSPL, both below and above the frequency of the incident field. The theory explains the Gibbs distribution of the anti-Stokes component of the NSPL spectrum, which is observed in our experiment. It also describes the increase in the integrated intensity of the anti-Stokes component of the spectrum with increasing metal temperature. In the Stokes part of the spectrum, the developed theory reproduces the plasmon resonance line when the plasmon resonance frequency lies below the frequency of the excitation field. In addition, the developed theory reproduces the shift of the maximum of the NSPL spectrum with respect to the plasmon resonant frequency.

NSPL accompanies SERS measurements manifesting itself in the scattering spectrum as a background. This background is superimposed on the SERS signal from the studied molecules, which limits the sensitivity of SERS. The proposed mechanism for NSPL and the microscopic theory of this phenomenon developed in this work shows that attempts to increase the SERS accuracy by redesigning substrates by reducing the size of the parts that form hot spots (points of the high field concentration) should also lead to an increase in the background, which degrades the properties of the substrates. This fact should be taken into account when designing substrates.

## ACKNOWLEDGEMENT

The authors thank I. A. Ryzhikov and I. A. Boginskaya for helpful discussions. The authors also thank Functional Micro/Nanosystem Laboratory from Bauman Moscow State Technical University and G. M. Arzumanyan and K. Mamatkulov for help in organizing and conducting experiments. A. A. L. acknowledges the support of the ONR under Grant No. N00014-20-1-2198. E. S. A. and V. Yu. S. thank the Foundation for the Advancement of Theoretical Physics and Mathematics "Basis."

# APPENDIX A. HAMILTONIAN OF THE SYSTEM

To quantitatively describe the spectrum of photoluminescence, we quantize the EM field of a plasmon nanoparticle at each frequency of the SSPNS. For this purpose, we use the procedure of macroscopic quantization of EM fields in a medium [19,20], which has been developed for linear media. The advantage of this approach is that the theory only relies on the general linear properties of the medium and does not require complicated first-principle calculations [31]. This procedure considers the medium as a set of elastic quantum dipoles representing oscillation states of electrons [31]. It is assumed that the dipoles interact with the modes of a reservoir. Such a reservoir can be phonons, impurities, etc. The eigenmodes of such a system are the coupled elastic dipoles and reservoir modes. The eigenmodes are found using the Fano diagonalization procedure [20,21]. The frequencies of these eigenmodes are real, and they differ from the plasmon resonance frequency. The closer the eigenmode frequency to the plasmon resonance frequency, the greater the contribution of the plasmon structure mode into this eigenmode.

At the same time, this procedure allows for obtaining the permittivity of the media. Since the coupling constant between the quantum dipoles with a reservoir is still a free parameter, it can be adjusted in a way that the obtained permittivity coincides with the values of the metal permittivity, $\varepsilon_M^{lin}$.

As a result of applying the macroscopic quantization procedure to a metal nanoparticle, it turns out that the eigenmode frequencies fill the SSPNS. The Hamiltonian of such eigenmodes has the form

$$\hat{H}_{pl} = \sum_k \hbar \omega_k \hat{a}_k^\dagger \hat{a}_k, \tag{A1}$$

Here $\hat{a}_k$ and $\hat{a}_k^\dagger$ are the operators of annihilation and creation of the eigenmode having the frequency $\omega_k$. The eigenmodes obtained after the Fano diagonalization are the quasiparticles considered in [20,21]. The eigenmode is a dressed state consisting of driven oscillations of the filed in the nanoparticle and the reservoir modes. Operators $\hat{a}_k$ and $\hat{a}_k^\dagger$ satisfy the Bose commutation relations:

$$\left[\hat{a}_k, \hat{a}_{k'}^\dagger\right] = \delta_{kk'}. \tag{A2}$$

The operator of the electric near field of these eigenmodes is expressed through the operators $\hat{a}_k$ and $\hat{a}_k^\dagger$ according to

$$\mathbf{E}(\mathbf{r},t) = \sum_k \mathbf{\Lambda}_k(\mathbf{r})\left(\hat{a}_k^\dagger(t) + \hat{a}_k(t)\right), \tag{A3}$$

where $\Lambda_k(\mathbf{r})$ are the functions that determine the distribution of the near-field created by the eigenmode with the frequency $\omega_k$. For example, for the case of a spherical subwavelength plasmon particle with the radius $R$ placed in a vacuum, the function $\Lambda_k(\mathbf{r})$ has the form :

$$\Lambda_k(\mathbf{r}) = \sqrt{\frac{\hbar \Delta \omega}{\pi R^3 \varepsilon_0}} \frac{\sqrt{\mathrm{Im}\,\varepsilon_M^{(lin)}(\omega_k)}}{\left|\varepsilon_M^{(lin)}(\omega_k)+2\right|} \mathbf{e}_d \begin{cases} R\,\mathrm{grad}\left[(r/R)\Psi(\theta,\varphi)\right], & r<R \\ R\,\mathrm{grad}\left[(R/r)^2 \Psi(\theta,\varphi)\right], & r>R \end{cases}$$

where $\varepsilon_0$ is the dielectric constant of vacuum, $\Delta \omega$ is the sampling frequency of the SSPNS, the radius-vector $\mathbf{r}$ is expressed in terms of polar, $\theta$, and azimuthal, $\varphi$, angles as $\mathbf{r} = [r\sin\theta\cos\varphi, r\sin\theta\sin\varphi, r\cos\theta]$, and $\mathbf{e}_d$ is the unit vector in the direction of the dipole moment. We emphasize once again that in the expression for the plasmon electric field operator, only the linear part of the permittivity $\varepsilon_M^{(lin)}(\omega_n)$ is used. Below, we use that the electric field per one quantum, $\Lambda_k(\mathbf{r})$, is inversely proportional to the volume of the nanoparticle, $\Lambda_k(\mathbf{r}) \sim 1/\sqrt{R^3} \sim 1/\sqrt{V}$.

Let us consider the interaction of the nanoparticle with an external monochromatic wave with the frequency $\omega_0$. Due to the monochromatic character of the external field, the excited quasiparticle has the frequency $\omega_0$. This interaction is described by the Hamiltonian $\hat{H}_{ex}$

$$\hat{H}_{ex} = -\hat{\mathbf{d}}_0 \cdot \mathbf{E}(\mathbf{r}_{pl})\cos(\omega_0 t) = \hbar\Omega_{ex}\hat{a}_0^\dagger e^{-i\omega_0 t} + \mathrm{h.c.}, \qquad (A4)$$

where $\hat{\mathbf{d}}_0 = \mathbf{d}_0(\hat{a}_0^\dagger + \hat{a}_0)$ is the operator of the dipole moment of a quasiparticle with the frequency $\omega_0$, $\Omega_{ex} = -\mathbf{d}_0 \cdot \mathbf{E}(\mathbf{r}_{pl})/\hbar$ is the Rabi frequency of the interaction of the quasiparticles with the external field. The value of the dipole moment $\mathbf{d}_0 \equiv \mathbf{d}(\omega_0)$ is determined by the geometry and permittivity of the nanoparticle.

All quasiparticle dipole moments have the same dependence on the frequency and the nanoparticle radius. For example, the dipole moment, $\mathbf{d}(\omega)$, of a subwavelength spherical plasmon nanoparticle of radius $R$ placed in a vacuum is determined by the expression [21]

$$\mathbf{d}(\omega) = 4\pi R^3 \sqrt{\frac{\Delta\omega\hbar\varepsilon_0}{\pi R^3}} \frac{\sqrt{\mathrm{Im}\,\varepsilon_M^{(lin)}(\omega)}}{\left|\varepsilon_M^{(lin)}(\omega)+2\right|} \mathbf{e}_d, \qquad (A5)$$

The radiated intensity of the dipole moment, Eq. (A5), is proportional to the square of this dipole moment. In the neighborhood of the plasmon resonance, $\omega \approx \omega_{pl}$, the intensity can be evaluated as $\sim\left((\omega-\omega_{pl})^2 + \gamma_{pl}^2\right)^{-2}$, where we denote $\gamma_{pl} = \mathrm{Im}\,\varepsilon_M^{(lin)}(\omega_{pl})/\left(\partial\mathrm{Re}\,\varepsilon_M^{(lin)}(\omega_{pl})/\partial\omega\right)$. This quantity has the meaning of the width of the SSPNS.

Next, we find the Hamiltonian of the interaction of quasiparticles with phonons in metal, which gives an additional nonlinear part of $\mathrm{Im}\,\varepsilon_\mathrm{M}(\omega_\mathrm{pl})$. The electric field inside the medium induces the polarization, which displaces electrons from the equilibrium positions. In turn, displaced electrons lead to a displacement of nuclei of the crystal lattice relative to their equilibrium positions. Displaced nuclei excite phonons, which result in the changing local density of the medium. This changing affects the dielectric constant of the metal [32]. Suppose that the nucleus at the point $\mathbf{r}$ deviates from the equilibrium position by $\mathbf{q}(\mathbf{r})$. We assume that this deviation is small enough [32] to expand the dielectric constant of the metal in the series over $\mathbf{q}(\mathbf{r})$. Retaining the zeroth and the first terms, we have:

$$\varepsilon(\mathbf{r},\omega) = \varepsilon^{(0)}(\mathbf{r},\omega) + w_i(\omega) q_i(\mathbf{r}), \tag{A6}$$

where the coefficients $w_i(\omega)$ determine the change in the dielectric constant of the metal at a point $\mathbf{r}$ due to the deviation of the nucleus. Substituting Eqs. (A6) and (A3) into the expression for the EM field energy $\frac{1}{8\pi}\int_V dV \left[\partial(\mathrm{Re}(\varepsilon)\omega)/\partial\omega\right]|\mathbf{E}(\mathbf{r})|^2$ and retaining the terms proportional to the first order of nucleus displacement, $q_i(\mathbf{r})$, we obtain the expression for the additional energy of the electric field:

$$\Delta\hat{W} = \frac{1}{8\pi}\int_V d^3\mathbf{r}\,\tilde{w}_i q_i(\mathbf{r}) \left(\mathbf{\Lambda}_n^{(0)*}(\mathbf{r})\hat{a}_n^\dagger + \mathbf{\Lambda}_n^{(0)}(\mathbf{r})\hat{a}_n\right)\left(\mathbf{\Lambda}_m^{(0)*}(\mathbf{r})\hat{a}_m^\dagger + \mathbf{\Lambda}_m^{(0)}(\mathbf{r})\hat{a}_m\right), \tag{A7}$$

where $\tilde{w}_i = \partial(\omega w_i)/\partial\omega$. For simplicity, we suppose that $\tilde{w}_i$ does depend on $i$, $\tilde{w}_i = \tilde{w}$.

According to the general theory of phonons in metals [32], the operator of the deviation of the $i$th nucleus from the equilibrium position, $\hat{q}_i$, can be expressed in terms of operators of annihilation, $\hat{b}_k$, and creation, $\hat{b}_k^\dagger$, of the $k$th phonon mode in the following way:

$$\hat{q}_i(\mathbf{r}) = \sum_l \left(C_l^*(\mathbf{r})\hat{b}_l^\dagger + C_l(\mathbf{r})\hat{b}_l\right), \tag{A8}$$

where operators $\hat{b}_l$ and $\hat{b}_l^\dagger$ obey the following commutation relations:

$$\left[\hat{b}_l, \hat{b}_{l'}^\dagger\right] = \delta_{l,l'}. \tag{A9}$$

The functions $C_l(\mathbf{r})$ describe the distribution of the $l$th phonon mode. They satisfy the normalization condition:

$$\int_V d^3\mathbf{r}\,|C_l(\mathbf{r})|^2 = 1 \tag{A10}$$

and their dynamics is described by the Hamiltonian

$$\hat{H}_\mathrm{phon} = \sum_l \hbar\omega_l^\mathrm{ph}\hat{b}_l^\dagger\hat{b}_l \tag{A11}$$

Substituting Eq. (A8) into Eq. (A7) and discarding the terms proportional to $\hat{a}_n^\dagger \hat{a}_{n'}^\dagger$ and $\hat{a}_n \hat{a}_{n'}$, we reduce the expression for the interaction Hamiltonian of plasmons and phonons to the optomechanical Hamiltonian [18]

$$\hat{H}_{\text{pl-phon}} = \sum_l \sum_{n,m} w_{nml} \left( C_l^*(\mathbf{r}) \hat{b}_l^\dagger + C_l(\mathbf{r}) \hat{b}_l \right) \hat{a}_n^\dagger \hat{a}_m, \qquad (A12)$$

where $w_{inml}$ is the interaction constant of phonons and plasmons having the form

$$w_{nml} = \tilde{w} \int_V d^3\mathbf{r} \, \Lambda_n^{(0)*}(\mathbf{r}) \Lambda_m^{(0)}(\mathbf{r}) C_l^*(\mathbf{r}). \qquad (A13)$$

Before proceeding further, we shortly describe the dependence of the interaction constant $w_{nml}$ between the EM field and phonons on the sample volume. As we mention above, $\Lambda_m^{(0)}(\mathbf{r}) \sim 1/\sqrt{V}$ and $C_l(\mathbf{r}) \sim 1/\sqrt{V}$. Assuming that $w$ depends only on the material properties, we obtain that $w_{nml} \sim 1/\sqrt{V}$. A different dependence takes place if we consider Brillouin scattering on the metal surface. Indeed, in this case, the field penetrates the metal within the skin depth $\delta \sim 10 \text{ nm}$. Then, in Eq. (A13), the integration volume is not equal to the total sample volume but is $V_{\text{int}} = A\delta$ where $A$ is the sample area, consequently, $w_{nml} \sim (\delta/L)/\sqrt{V}$ where $L$ is the thickness of the sample. Taking into account that phonons have a finite coherence length, we should replace the total sample thickness $L$ by the coherence length $L_{\text{coh}} \sim 10 \, \mu\text{m}$. As a result, the interaction between the phonons and the EM field in the case of Brillouin scattering on a metal surface differs by the factor $\delta/L_{\text{coh}} \sim 10^{-3}$ from Brillouin scattering in metal nanoparticles. This qualitative estimation can explain why the Brillouin scattering background on a metal nanoparticle is much larger than that on the metal surface.

To describe radiation of dipole eigenmodes, it is necessary to consider their interaction with free space modes. For this purpose, the Hamiltonian of the EM field of a free space mode is usually introduced:

$$\hat{H}_{\text{rad}} = \sum_{\mu=1,2} \int d^3\mathbf{k} \, \hbar c k \hat{f}^\dagger(\mu, \mathbf{k}) \hat{f}(\mu, \mathbf{k}). \qquad (A14)$$

Here $\hat{f}(\mu, \mathbf{k})$ and $\hat{f}^\dagger(\mu, \mathbf{k})$ are annihilation and creation operators of the EM field of the free space mode with the polarization $\mu$ and the wave vector $\mathbf{k}$, satisfying the following commutation relations:

$$\left[ \hat{f}(\mu, \mathbf{k}), \hat{f}^\dagger(\mu', \mathbf{k}') \right] = \delta_{\mu\mu'} \delta(\mathbf{k} - \mathbf{k}'). \qquad (A15)$$

The interaction between the dipole moments of the eigenmodes of a nanoparticle and the electric field of the free space modes has the form

$$\hat{H}_{\text{pl-rad}} = \sum_n -\hat{\mathbf{d}}_n \cdot \hat{\mathbf{E}}_n = \sum_{\mu=1,2} \int d^3\mathbf{k} \sum_n \hbar \Omega_n(\mu, \mathbf{k}) \hat{a}_n^\dagger \hat{f}(\mu, \mathbf{k}) + \text{h.c.} \qquad (A16)$$

Here $\Omega_n(\mu,\mathbf{k}) = -\mathbf{d}_n \cdot \mathbf{E}_{\mu,\mathbf{k}}(\mathbf{r})/\hbar$ is the Rabi constant of the interaction of the electric field of the free space mode with the polarization $\mu$ and the wave vector $\mathbf{k}$ with the eigenmode of the plasmon particle.

Thus, the described system includes plasmons of a subwavelength metal nanoparticle, the external EM wave incident on the nanoparticle, free space photons, and phonons of the metal from which the nanoparticle is made. The Hamiltonian of this system is the sum of the Hamiltonians (A1), (A4), (A11)–(A14), and (A16):

$$\hat{H} = \hat{H}_{pl} + \hat{H}_{phot} + \hat{H}_{pl\text{-}phot} + \hat{H}_{ex} + \hat{H}_{phon} + \hat{H}_{pl\text{-}phon}. \tag{A17}$$

## APPENDIX B. ELIMINATION OF PHONON DEGREES OF FREEDOM IN THE MARKOV APPROXIMATION

Due to a large number of degrees of freedom described by Hamiltonian (A17), it is difficult to find the system eigenmodes even numerically. We, therefore, should make some simplifications. We exclude the degrees of freedom of phonons in metal and free-space photons. After this, we have to deal with an open quantum system, which only includes the plasmon-polariton modes, instead of considering a closed quantum system, which includes photons, phonons, and EM modes of a sphere excited by an external field. Consequently, it is more convenient to use the master equation for the density matrix instead of the Heisenberg equation.

To obtain the master equation, it is necessary to exclude the photon and phonon subsystems sequentially. Employing the standard procedure, we obtain the Lindblad superoperator $\hat{L}_{photon}(\hat{\rho})$, describing the relaxation of the plasmon density matrix $\hat{\rho}$ [27]:

$$\hat{L}_{photon}(\hat{\rho}) = \sum_n \gamma_{rad}(\omega_n)\{2\hat{a}_n \hat{\rho} \hat{a}_n^\dagger - \hat{\rho}\hat{a}_n^\dagger \hat{a}_n - \hat{a}_n^\dagger \hat{a}_n \hat{\rho}\}, \tag{B1}$$

where the decay rate $\gamma_{rad}(\omega_n)$ of the plasmon-polariton at the frequency $\omega_n$ is determined according to the Fermi's golden rule [27]:

$$\gamma_{rad}(\omega_n) = \pi \sum_{\mu=1,2} \int d^3\mathbf{k} \, |\Omega_n(\mu,\mathbf{k})|^2 \, \delta(\omega_n - c|\mathbf{k}|). \tag{B2}$$

The rate of plasmon-polariton energy loss at the frequency $\omega_n$ is determined by the dipole moment of the transition of this mode and is proportional to $\gamma_{rad}(\omega_n) \propto 1/|\varepsilon_M(\omega_n)+2|^2$ [21]. The Lindblad superoperator (B1) describes the process of plasmon energy loss due to radiation into free space [33]. As a result, the term corresponding to the free space modes is excluded from the Hamiltonian of the whole closed system, but Lindbladian (B1) appears in the master equation:

$$\frac{\partial \hat{\rho}}{\partial t} = \frac{i}{\hbar}\left[\hat{\rho}, \hat{H}_{pl} + \hat{H}_{ex} + \hat{H}_{phon} + \hat{H}_{pl\text{-}phon}\right] + \hat{L}_{photon}[\hat{\rho}]. \tag{B3}$$

Now, we can exclude the phonon subsystem. For this purpose, we assume that at any moment of time, the phonons are in a state of thermodynamic equilibrium with a fixed temperature $T$. For convenience, in Hamiltonian (A12), we transition to the interaction representation:

$$\hat{H}_{\text{pl-phon}} = \sum_{l}\sum_{n,m} w_{nml} \left( C_l^*(\mathbf{r}) \hat{b}_l^\dagger e^{i\omega_l^{\text{ph}} t} + C_l(\mathbf{r}) \hat{b}_l e^{-i\omega_l^{\text{ph}} t} \right) \hat{a}_n^\dagger e^{i\omega_n t} \hat{a}_m e^{-i\omega_m t}. \quad (B4)$$

The Hamiltonian of the interaction of phonons and plasmons, Eq. (B4), describes the inelastic process of absorption/emission of a phonon with the energy transfer from the plasmon-polariton with the frequency $\omega_n$ to the one with the frequency $\omega_m$. This is the process of Brillouin scattering.

Following the standard procedure of the reservoir exclusion [27], in which the phonon subsystem with Hamiltonian (A11) serves as a reservoir, we obtain the following Lindblad relaxation superoperator $\hat{L}_{\text{phonons}}(\hat{\rho})$ describing the relaxation of the density matrix of the plasmon subsystem $\hat{\rho}$

$$\hat{L}_{\text{phonon}}(\hat{\rho}) = \sum_n \sum_m G_{\omega_m,\omega_n} \left\{ 2\hat{a}_n^\dagger \hat{a}_m \hat{\rho} \hat{a}_m^\dagger \hat{a}_n - \hat{\rho} \hat{a}_m^\dagger \hat{a}_n \hat{a}_n^\dagger \hat{a}_m - \hat{a}_m^\dagger \hat{a}_n \hat{a}_n^\dagger \hat{a}_m \hat{\rho} \right\}. \quad (B5)$$

In Eq. (B5), the function $G(\omega_n - \omega_m)$ is the rate of the relaxation processes, which is determined from the Fermi's golden rule [27]

$$G_{\omega_m,\omega_n} = \begin{cases} \pi \sum_l |w_{nml}|^2 \left(1 + N(\omega_l^{\text{ph}})\right) \delta(\omega_n - \omega_m - \omega_l^{\text{ph}}), & \omega_n - \omega_m > 0, \\ \pi \sum_l |w_{nml}|^2 N(\omega_l^{\text{ph}}) \delta(\omega_m - \omega_n - \omega_l^{\text{ph}}), & \omega_n - \omega_m < 0. \end{cases} \quad (B6)$$

After substitution of Eq. (6) to Eq. (B6), we obtain Eq. (7) from the main text.

The function $G_{\omega_m,\omega_n}$ has two important features. First, since above the frequency $\omega_{\max}^{\text{ph}}$, the density of states of phonons is zero, then

$$G_{\omega_m,\omega_n} = 0, \quad \text{when} \quad \omega_m - \omega_n > \omega_{\max}^{\text{ph}}. \quad (B7)$$

Second, since phonons are in thermodynamic equilibrium with temperature $T$, one obtains [27]

$$G_{\omega_m,\omega_n} \propto \exp(-\hbar(\omega_m - \omega_n)/k_B T). \quad (B8)$$

In Eq. (B6), using the phonon density of states $D(\omega^{\text{ph}})$, the sum over phonon wavevectors $\mathbf{k}$ can be transformed into the integral over frequencies $\omega_l^{\text{ph}}$ as $\sum_l = \int d\omega_{\text{ph}} D(\omega_{\text{ph}})$. To advance further, we assume that the crystal lattice is isotropic, i.e., the constant $w_{inml}$ does not depend on $i$, and in the frequency range from 0 to $\omega_{\max}^{\text{ph}}$, phonons have linear dispersion, $\omega_{\text{ph}} = u|\mathbf{k}|$, where $u$ is the sound speed. In this case, for the density of states we have $D(\omega_{\text{ph}}) = 3V\omega_{\text{ph}}^2 / 2\pi^2 u^3$ [34]. Then, performing the integration in Eq. (B6) and using the expression for the function $w_{nml}$, Eq. (8) from the main text, we obtain

$$G_{\omega_m,\omega_n} = \frac{w^2\hbar^3\omega_n\omega_m}{2\pi Mc_{ph}^3}(\omega_n-\omega_m)(1+N(\omega_n-\omega_m))$$

$$\times\left(\sin\left(\frac{(\omega_n-\omega_m)R}{v_s}\right)-\frac{(\omega_n-\omega_m)R}{v_s}\cos\left(\frac{(\omega_n-\omega_m)R}{v_s}\right)\right)^2 (Vk_l^3)^{-2}, \quad 0\leq\omega_n-\omega_m\leq\omega_{ph}^{max}, \tag{B9}$$

In numerical simulation, we use the parameter $w$ as a fitting parameter.

As a result, we obtain the master equation, from which phonon Hamiltonians (A11) and (B4) are excluded, and which contains a new dissipative operator $\hat{L}_{phonon}(\hat{\rho})$:

$$\frac{\partial\hat{\rho}}{\partial t} = \frac{i}{\hbar}\left[\hat{\rho},\hat{H}_{pl}+\hat{H}_{ex}\right]+\hat{L}_{photon}(\hat{\rho}). \tag{B10}$$

The relaxation operator (B5) reflects the structure of excluded Hamiltonian (A12) and describes the process of Brillouin scattering, in which the quasiparticle with the frequency $\omega_n$ turns into the quasiparticle with the frequency $\omega_m$ due to the scattering on the phonon with the frequency $\omega_n-\omega_m$. If $\omega_n-\omega_m>0$, then the process corresponds to the Stokes scattering, and the energy is absorbed by the phonon reservoir. In the opposite case, $\omega_n-\omega_m<0$ this is the anti-Stokes scattering, in which the energy of a quasiparticle increases by $\hbar(\omega_m-\omega_n)$. This energy is drawn from the thermal energy of the phonon reservoir. From Eq. (B8) it follows that the speed of the first process is greater than the speed of the second one by the factor of $(1+N(\omega_{\mathbf{k}}^{ph}))/N(\omega_{\mathbf{k}}^{ph})=\exp(\hbar\Delta\omega/k_B T)$, so that we have a more intense excitation of low-frequency modes.

Finally, using the identities $\langle\dot{\hat{a}}_k\rangle=\text{Tr}(\dot{\hat{\rho}}\hat{a}_k)\equiv a_k$ and $\langle\dot{\hat{n}}_k\rangle=\text{Tr}(\dot{\hat{\rho}}\hat{n}_k)=n_k$, commutation relation (A2), and the master equation for the density matrix, Eq. (B10), we arrive at the equation for the expected value of operators $\hat{a}_k$ and $\hat{n}_k$. For $a_k$ we obtain

$$\frac{da_k}{dt} = \frac{i}{\hbar}\text{Tr}\left(\hat{a}_k\left[\sum_n\hbar\omega_n\hat{a}_n^\dagger\hat{a}+\hbar\Omega_{ex}\hat{a}_0^\dagger e^{-i\omega_0 t}+\hbar\Omega_{ex}\hat{a}_0 e^{-i\omega_0 t},\hat{\rho}\right]\right)$$

$$+\text{Tr}\left(a_k\sum_n\sum_m G_{\omega_m,\omega_n}\left\{2\hat{a}_n^\dagger\hat{a}_m\hat{\rho}\hat{a}_m^\dagger\hat{a}_n-\hat{\rho}\hat{a}_m^\dagger\hat{a}_n\hat{a}_n^\dagger\hat{a}_m-\hat{a}_m^\dagger\hat{a}_n\hat{a}_n^\dagger\hat{a}_m\hat{\rho}\right\}\right)$$

$$+\text{Tr}\left(a_k\sum_n\gamma_{rad}(\omega_n)\left\{2\hat{a}_n\hat{\rho}\hat{a}_n^\dagger-\hat{\rho}\hat{a}_n^\dagger\hat{a}_n-\hat{a}_n^\dagger\hat{a}_n\hat{\rho}\right\}\right)$$

$$=(-i\omega_k)\text{Tr}(\hat{a}_k\hat{\rho})-i\Omega_{ex}e^{-i\omega_0 t}\delta_{k,0}\text{Tr}(\hat{\rho})+\sum_m G_{\omega_m,\omega_k}\text{Tr}(\hat{a}_k\hat{n}_m\hat{\rho})-G_{\omega_k,\omega_m}\text{Tr}(\hat{a}_k(\hat{n}_m+1)\hat{\rho}) \tag{B11}$$

$$-G_{\omega_k,\omega_k}\text{Tr}(\hat{a}_k\hat{\rho})-\gamma_{rad}(\omega_k)\text{Tr}(\hat{a}_k\hat{\rho})$$

$$=\left(-i\omega_k-\gamma_{rad}(\omega_k)-G_{\omega_k,\omega_k}\right)a_k-i\Omega_{ex}e^{-i\omega_0 t}\delta_{k,0}$$

$$+\sum_m\left\{G_{\omega_m,\omega_k}\langle\hat{a}_k\hat{n}_m\rangle-G_{\omega_k,\omega_m}\left[\langle\hat{a}_k\hat{n}_m\rangle+\langle\hat{a}_k\rangle\right]\right\}.$$

Further, we take into account that $G_{\omega_k,\omega_k} = 0$ [see Eq. (B9)], and in the mean-field approximation, we uncouple the correlations $\langle \hat{a}_k \hat{n}_m \rangle = \langle \hat{a}_k \rangle \langle \hat{n}_m \rangle$. As a result, we have:

$$\frac{da_k}{dt} = \left(-i\omega_k - \gamma_{\rm rad}(\omega_k)\right)a_k - i\Omega_{\rm ex} e^{-i\omega_0 t}\delta_{k,0} + a_k \sum_m \left\{ G_{\omega_m,\omega_k} n_m - G_{\omega_k,\omega_m}(n_m + 1)\right\}. \quad \text{(B12)}$$

These are Eqs. (2)–(3) from the main text. Similarly, we obtain Eq. (4) for $n_k$, $k \neq 0$:

$$\frac{dn_k}{dt} = \frac{i}{\hbar}{\rm Tr}\left(n_k\left[\sum_n \hbar\omega_n \hat{a}_n^\dagger \hat{a} + \hbar\Omega_{\rm ex}\hat{a}_0^\dagger e^{-i\omega_0 t} + \hbar\Omega_{\rm ex}\hat{a}_0^\dagger e^{-i\omega_0 t}, \hat{\rho}\right]\right)$$

$$+ {\rm Tr}\left(n_k \sum_n \sum_m G_{\omega_m,\omega_n}\left\{2\hat{a}_n^\dagger \hat{a}_m \hat{\rho} \hat{a}_m^\dagger \hat{a}_n - \hat{\rho}\hat{a}_m^\dagger \hat{a}_n \hat{a}_n^\dagger \hat{a}_m - \hat{a}_m^\dagger \hat{a}_n \hat{a}_n^\dagger \hat{a}_m \hat{\rho}\right\}\right)$$

$$+ {\rm Tr}\left(n_k \sum_n \gamma_{\rm rad}(\omega_n)\left\{2\hat{a}_n \hat{\rho}\hat{a}_n^\dagger - \hat{\rho}\hat{a}_n^\dagger \hat{a}_n - \hat{a}_n^\dagger \hat{a}_n \hat{\rho}\right\}\right) \quad \text{(B13)}$$

$$= \sum_m G_{\omega_m,\omega_k}{\rm Tr}\left((1+\hat{n}_k)\hat{n}_m\hat{\rho}\right) - G_{\omega_k,\omega_m}{\rm Tr}\left(\hat{n}_k(\hat{n}_m+1)\hat{\rho}\right) - 2\gamma_{\rm rad}(\omega_k){\rm Tr}(\hat{n}_k\hat{\rho})$$

$$\approx -2\gamma_{\rm rad}(\omega_k)n_k + 2\sum_m \left\{ G_{\omega_m,\omega_k}n_m[1+n_k] - G_{\omega_k,\omega_m}n_k[1+n_m]\right\}.$$

**References**

[1] A. Mooradian, *Photoluminescence of metals*, Phys. Rev. Lett. **22**, 185 (1969).
[2] G. T. Boyd, Z. H. Yu, and Y. R. Shen, *Photoinduced luminescence from the noble metals and its enhancement on roughened surfaces*, Phys. Rev. B **33**, 7923 (1986).
[3] M. B. Mohamed, V. Volkov, S. Link, and M. A. El-Sayed, *The'lightning'gold nanorods: fluorescence enhancement of over a million compared to the gold metal*, Chem. Phys. Lett. **317**, 517 (2000).
[4] O. P. Varnavski, M. B. Mohamed, M. A. El-Sayed, and T. Goodson, *Relative enhancement of ultrafast emission in gold nanorods*, J. Phys. Chem. B **107**, 3101 (2003).
[5] P. Apell, R. Monreal, and S. Lundqvist, *Photoluminescence of noble metals*, Phys. Scr. **38**, 174 (1988).
[6] M. R. Beversluis, A. Bouhelier, and L. Novotny, *Continuum generation from single gold nanostructures through near-field mediated intraband transitions*, Phys. Rev. B **68**, 115433 (2003).
[7] E. Dulkeith, T. Niedereichholz, T. A. Klar, J. Feldmann, G. Von Plessen, D. I. Gittins, K. S. Mayya, and F. Caruso, *Plasmon emission in photoexcited gold nanoparticles*, Phys. Rev. B **70**, 205424 (2004).
[8] K.-Q. Lin, J. Yi, J.-H. Zhong, S. Hu, B.-J. Liu, J.-Y. Liu, C. Zong, Z.-C. Lei, X. Wang, and J. Aizpurua, *Plasmonic photoluminescence for recovering native chemical information from surface-enhanced Raman scattering*, Nat. Commun. **8**, 14891 (2017).
[9] J. T. Hugall and J. J. Baumberg, *Demonstrating photoluminescence from Au is electronic inelastic light scattering of a plasmonic metal: the origin of SERS backgrounds*, Nano Lett. **15**, 2600 (2015).

[10] Y. He, J.-J. Li, and K.-D. Zhu, *A tunable optical response of a hybrid semiconductor quantum dot-metal nanoparticle complex in the presence of optical excitations*, J. Opt. Soc. Am. B **29**, 997 (2012).

[11] D. A. Weitz, T. J. Gramila, A. Z. Genack, and J. I. Gersten, *Anomalous low-frequency Raman scattering from rough metal surfaces and the origin of surface-enhanced Raman scattering*, Phys. Rev. Lett. **45**, 355 (1980).

[12] S. Garoff, D. A. Weitz, T. J. Gramila, and C. D. Hanson, *Optical absorption resonances of dye-coated silver-island films*, Opt. Lett. **6**, 245 (1981).

[13] D. A. Weitz, S. Garoff, C. D. Hanson, T. J. Gramila, and J. I. Gersten, *Fluorescent lifetimes of molecules on silver-island films*, Opt. Lett. **7**, 89 (1982).

[14] J. Kabuss, A. Carmele, T. Brandes, and A. Knorr, *Optically driven quantum dots as source of coherent cavity phonons: a proposal for a phonon laser scheme*, Phys. Rev. Lett. **109**, 054301 (2012).

[15] V. Y. Shishkov, E. S. Andrianov, A. A. Pukhov, A. P. Vinogradov, and A. A. Lisyansky, *Enhancement of the Raman Effect by Infrared Pumping*, Phys. Rev. Lett. **122**, 153905 (2019).

[16] M. Reitz, C. Sommer, and C. Genes, *Langevin approach to quantum optics with molecules*, Phys. Rev. Lett. **122**, 203602 (2019).

[17] V. I. Fabelinsky, D. N. Kozlov, S. N. Orlov, Y. N. Polivanov, I. A. Shcherbakov, V. V. Smirnov, K. A. Vereschagin, G. M. Arzumanyan, K. Z. Mamatkulov, and K. N. Afanasiev, *Surface-enhanced micro-CARS mapping of a nanostructured cerium dioxide/aluminum film surface with gold nanoparticle-bound organic molecules*, J. Raman Spectrosc. **49**, 1145 (2018).

[18] A. Yariv, *Quantum electronics* 3rd ed. (Wiley, New York, 1989).

[19] B. Huttner and S. M. Barnett, *Quantization of the electromagnetic field in dielectrics*, Phys. Rev. A **46**, 4306 (1992).

[20] T. G. Philbin, *Canonical quantization of macroscopic electromagnetism*, New J. Phys. **12**, 123008 (2010).

[21] V. Y. Shishkov, E. S. Andrianov, A. A. Pukhov, and A. P. Vinogradov, *Hermitian description of localized plasmons in dispersive dissipative subwavelength spherical nanostructures*, Phys. Rev. B **94**, 235443 (2016).

[22] B. Bennett, A. Maradudin, and L. Swanson, *A theory of the Brillouin scattering of light by acoustic phonons in a metal*, Ann. Phys. **71**, 357 (1972).

[23] A. Marvin, V. Bortolani, and F. Nizzoli, *Surface Brillouin scattering from acoustic phonons. I. General theory*, J. Phys. C Solid State **13**, 299 (1980).

[24] A. Penzkofer, A. Laubereau, and W. Kaiser, *High intensity Raman interactions*, Progr. Quant. Electron. **6**, 55 (1979).

[25] L. F. Mollenauer, J. C. White, and C. R. Pollock, *Tunable lasers* (Springer, 1992).

[26] A. Lombardi, M. K. Schmidt, L. Weller, W. M. Deacon, F. Benz, B. de Nijs, J. Aizpurua, and J. J. Baumberg, *Pulsed molecular optomechanics in plasmonic nanocavities: from nonlinear vibrational instabilities to bond-breaking*, Phys. Rev. X **8**, 011016 (2018).

[27] H.-P. Breuer and F. Petruccione, *The theory of open quantum systems* (Oxford University Press, New York, 2002).

[28] M. O. Scully and M. S. Zubairy, *Quantum optics* (AAPT, College Park, Maryland, 1999).

[29] W. Drexel, W. Gläser, and F. Gompf, *Phonon dispersion in silver*, Phys. Lett. A **28**, 531 (1969).

[30]  J. Lynn, H. Smith, and R. Nicklow, *Lattice dynamics of gold*, Phys. Rev. B **8**, 3493 (1973).
[31]  W. Vogel and D.-G. Welsch, *Quantum optics* (Wiley, Weinheim, 2006).
[32]  R. P. Feynman, *Statistical Mechanics* (Benjamin, California, 1972).
[33]  Y. E. Lozovik, I. A. Nechepurenko, E. S. Andrianov, A. V. Dorofeenko, A. A. Pukhov, and N. M. Chtchelkatchev, *Self-consistent description of graphene quantum amplifier*, Phys. Rev. B **94**, 035406 (2016).
[34]  L. D. Landau and E. M. Lifshitz, *Statistical Physics, Part 1: V. 5* (Pergamon Press, Fairview Park, New York, 1980).